\documentclass[reprint, superscriptaddress, showpacs,amsmath,amssymb, aps, prb]{revtex4-2}
\usepackage{graphicx} 
\usepackage{subfigure}
\usepackage{xcolor} 
\usepackage[english]{babel}
\usepackage[utf8]{inputenc}
\usepackage{url}
\usepackage{dcolumn}
\usepackage{bm}
\usepackage[colorlinks=true,allcolors=blue]{hyperref}

\providecommand{\vect}[1]{{\boldsymbol{#1}}}

\begin{document}

\title{Topological Classification of Non-Normalizable Vector Fields}%

\author{Philipp Gessler}
\affiliation{Faculty of Physics and Center for Nanointegration Duisburg-Essen (CENIDE), University of Duisburg-Essen, 47048 Duisburg, Germany}

\author{Alessandro Pignedoli}
\affiliation{Faculty of Physics and Center for Nanointegration Duisburg-Essen (CENIDE), University of Duisburg-Essen, 47048 Duisburg, Germany}

\author{Alexander Neuhaus}
\affiliation{Faculty of Physics and Center for Nanointegration Duisburg-Essen (CENIDE), University of Duisburg-Essen, 47048 Duisburg, Germany}

\author{Frank-J.~Meyer zu Heringdorf}
\affiliation{Faculty of Physics and Center for Nanointegration Duisburg-Essen (CENIDE), University of Duisburg-Essen, 47048 Duisburg, Germany}

\author{Maria Azhar}
\affiliation{Faculty of Physics and Center for Nanointegration Duisburg-Essen (CENIDE), University of Duisburg-Essen, 47048 Duisburg, Germany}

\author{Karin Everschor-Sitte}%
\affiliation{Faculty of Physics and Center for Nanointegration Duisburg-Essen (CENIDE), University of Duisburg-Essen, 47048 Duisburg, Germany}

\date{\today} 

\begin{abstract}
Topological classification of physical vector fields conventionally relies on field normalization and homotopy-based invariants. However, when field amplitudes vanish, normalization becomes ill-defined, preventing a direct topological characterization.
Here, we introduce a general framework for the topological classification of non-normalizable $n$-dimensional vector fields with compactifiable base spaces by transforming them into $(n+1)$-dimensional normalized vector fields. This construction extends homotopy-based classification to fields containing amplitude zeros. We explicitly demonstrate the approach for one-, two-, and three-dimensional non-normalized vector fields and derive the corresponding topological invariants.  
The resulting topological charges are robust under continuous deformations and can change only when the embedding structure becomes singular. 
Our framework provides a unified route to the topological characterization of non-normalizable fields and opens the door to the study of topological phenomena in a broad range of systems, including magnetic textures, ferroelectrics, electromagnetic fields, and wave systems.
\end{abstract}

\maketitle

\section{Introduction}
Topology has become a central organizing principle in modern physics, providing a robust framework for classifying states and textures beyond local geometric details\cite{Wen2017,Nakahara2003}.
Topological invariants, such as winding numbers, enable the distinction of configurations that cannot be continuously deformed into one another without crossing a singularity~\cite{Mermin1979}. This perspective has  been proven to be successful across diverse physical systems, ranging from magnetic textures~\cite{Bogdanov1994,Muehlbauer2009,Azhar2026,koraltan2026} 
and liquid crystals~\cite{DeGennes1993, Fukuda2011, Wu2022, Hall2025}
to ferroelectrics~\cite{Naumov2004, Yadav2016, Wang2023, Lukyanchuk2025}, superconductors~\cite{Abrikosov1956, Bailin1978, Kitaev_2001}, and wave phenomena~\cite{Delplace2017, Wang2025, Tsesses2018, Tsesses2025, Neuhaus2026, Smirnova2024, Shen2025}. The appeal of topological classification lies in its universality: configurations sharing the same topological invariant exhibit common global properties independent of microscopic details~\cite{Nakahara2003}.

Conventional topological classification of vector fields relies on normalized order parameters. For a continuous physical vector field $\vect{\tilde{v}}:\mathbb{R}^m \rightarrow \mathbb{R}^n$, the normalized field $\vect{\tilde{v}}/|\vect{\tilde{v}}|$ defines a mapping from base space $\mathbb{R}^m$ to a target space $S^{n-1}$ which is also referred to as the order parameter space in physical contexts.
Under the assumption that the base space domain can be compactified to $S^m$ (e.g.\ the field approaches a constant configuration at spatial infinity or satisfies compact boundary conditions and is simply connected), topological classes are determined by the associated homotopy group $\pi_m(S^{n-1})$.
Examples include topological vortices classified by the first homotopy group $\pi_1(S^1)$, and skyrmionic textures classified by $\pi_2(S^2)$~\cite{Abrikosov1956, Bogdanov1994, Mermin1979, Fert2017, Wang2022}.

However, this approach breaks down whenever the field amplitude vanishes. At zeros of the vector field, normalization becomes ill-defined, and the associated mapping to the target space is singular. Such amplitude zeros occur naturally in many physical systems, including polarization textures in ferroelectrics~\cite{Li2025, Yadav2016}, nodal structures in wave systems~\cite{Davis2020}, 
superconducting vortices characterized by vanishing order parameters at their cores~\cite{Abrikosov1956}, 
and magnetic configurations with Bloch points~\cite{DaCol2014, Azhar2022, Kuchkin2025, Kuchkin2026}. Consequently, a large class of physically relevant vector fields remains inaccessible to conventional topological characterization. Existing approaches often either exclude singular regions or treat them as defects~\cite{Mermin1979, Louis1980, Volovik2009}, thereby limiting a unified classification of the full field.

In this work, we introduce a general framework for the topological classification of vector fields with amplitude zeros, thereby extending topological analysis to vanishing fields. 
Our approach embeds the $n$-dimensional non-normalizable vector field into an $(n+1)$-dimensional space, resulting in a normalized vector field by construction and thereby enabling the systematic definition of topological invariants.
We explicitly demonstrate the method for one-, two-, and three-dimensional vector fields and show that it yields robust topological invariants distinguishing topologically inequivalent field configurations.

This paper is organized as follows. In Sec.~\ref{sec:Generalframework}, we introduce the general embedding procedure and establish the notation for field and target spaces. Sec.~\ref{sec:examples} presents explicit constructions and examples for one-, two-, and three-dimensional vector fields. Finally, in Sec.~\ref{sec:discon}, we discuss implications and potential applications to physical systems with singular or vanishing order parameters.

\section{General Framework for the Topological Classification of Non-Normalizable Vector Fields}~\label{sec:Generalframework}

We introduce a general procedure for the topological classification of vanishing continuous vector fields $\vect{\tilde{v}}$ by introducing a normalized higher-dimensional representation $\vect{V}$. Therefore, $\vect{V}$ admits a topological classification using homotopy group theory~\cite{Nakahara2003}.

We consider a vector field 
\begin{equation}
    \vect{\tilde{v}}:\mathbb{\bar{R}}^m\rightarrow \tilde{D}^n,
\end{equation}
 where $\mathbb{\bar{R}}^m$ is the compactified $m$-dimensional Euclidean space and $\tilde{D}^n$ is the target-space image of $\vect{\tilde{v}}$.
Here, we assume $\tilde{D}^n$ to be a closed, star-body with respect to the origin. We can parametrize $\tilde{D}^n$ using $n$-dimensional spherical coordinates with radial coordinate $\rho$ and angular coordinates $\chi_i$, where $i \in \{1,\ldots,n-1\}$. The boundary $\partial\tilde{D}^n$ is then described by the radial function $R(\chi_i)= \mathrm{max}_{\rho} |\tilde{\vect v}(\vect \chi_i)|$. 
The construction of $\vect{V}$ proceeds in two steps.

First, the vector field $\tilde{\vect{v}}$ is continuously rescaled such that the image of the resulting field $\vect{v}$ becomes the $n$-dimensional unit disk $D^n$.
To this end, the field magnitude is normalized independently along each angular direction $\chi_i$ by $R(\chi_i)$: 
\begin{subequations}
\label{eq:rescaling}
\begin{align}
\vect v &: \mathbb{R}^m \rightarrow D^n \subset \mathbb{R}^n, \\
\vect v(\rho,\chi_i) &= \frac{\tilde{\vect v}(\rho, \chi_i)}{R(\chi_i)} f(\tilde{\vect{v}}(\rho,\chi_i)).
\end{align}
\end{subequations}
Thus, each radial direction is rescaled to unit length, continuously deforming the original target space $\tilde D^n$ into the unit disk $D^n$.
The function $f(\tilde{\vect{v}})$ specifies how points in the target space are rescaled while leaving the boundary of $D^n$ fixed, thereby determining the real-space profile of $\vect{v}(\vect{r})$.
By construction, the rescaled field satisfies $|\vect{v}| \leq 1$, and the mapping is surjective, such that the image of $\vect{v}$ fully covers the unit disk $D^n$.

In the second step, the rescaled vector field is embedded in one higher dimension by introducing a normalized vector field $\vect{V}$.
\begin{subequations}
\label{eq:HigherMap}
\begin{align}
\vect{V}
&:\mathbb{R}^m \rightarrow S^n \subset \mathbb{R}^{n+1},\\
\vect{V}(\vect{r}) 
&=
\begin{pmatrix}
\vect{v}(\vect r)\\
\Gamma(\vect{v}(\vect r)) \, \sqrt{1-|\vect{v(\vect r)}|^2}
\end{pmatrix},
\end{align}
\end{subequations}
where the length of the vector field $\vect{v}$ is basically incorporated as another component of the vector. 
$\Gamma(\vect{v})$ is a scalar function satisfying 
$|\Gamma(\vect{v})|=1$, chosen such that the resulting field $\vect{V}$ remains continuous.
The choice of $\Gamma(\vect{v})$ is generally not unique. This non-uniqueness has a geometric origin and reflects the freedom in choosing a lift of the mapping into the higher-dimensional sphere, associated with the double-cover structure induced by the embedding (see App.~\ref{sec:choiceOfLift} for details).
By construction, $|\vect{V}|=1$, such that $\vect{V}$ defines a normalized mapping onto the sphere $S^n\subset \mathbb{R}^{n+1}$. 

The embedding procedure preserves information about both the orientation and amplitude of the original vector field while regularizing singularities associated with amplitude zeros.
Furthermore, the compactifiability of the original field $\tilde{\vect{v}}$ is preserved under the successive mappings to $\vect{v}$ and $\vect{V}$. Consequently, whenever $\tilde{\vect{v}}$ admits a compactification of the base space to $S^m$, the normalized field $\vect{V}$ defines a continuous mapping $\vect{V}: S^m \rightarrow S^n$, allowing for a topological classification in terms of the corresponding homotopy group $\pi_m(S^n)$. The present construction therefore provides a general framework for assigning topological invariants to vector fields with a compactifiable base space that are non-normalizable in the conventional sense.

In the following, we illustrate the general embedding procedure through explicit examples in one, two, and three dimensions. These examples demonstrate the construction for specific vector fields and highlight physical systems in which such configurations naturally arise.

\section{Explicit Examples of the Embedding Procedure}
\label{sec:examples}

Here we present the application of the previously described general procedure, showing specific examples representative of the most common situation in physics.

\begin{figure*}[tb]
    \centering
    \includegraphics[width=\textwidth]{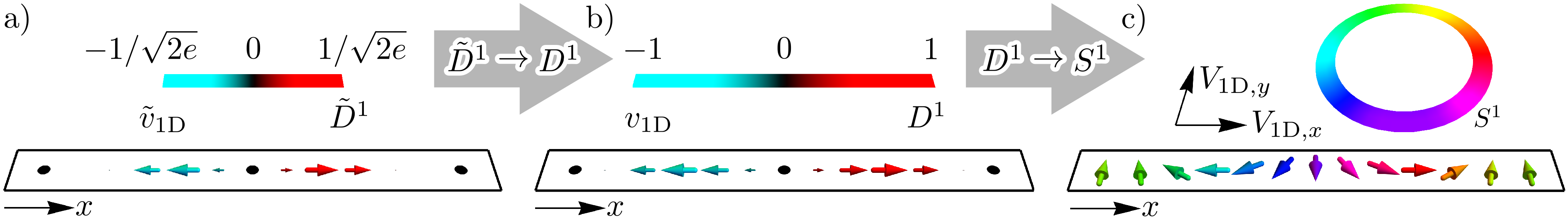}
    \caption{
      Illustration of the mapping of a 1D field $\tilde{v}_{\mathrm{1D}}(x)$ from $\tilde{D}^1$ into a field $v_\mathrm{1D}$ with an order parameter on $D^1$, followed by the mapping to a higher dimensional vector $V_\mathrm{1D}$ with an order parameter $S^1\subset\mathbb{R}^2$.
      In each panel, the order-parameter space is shown above the corresponding real-space field.
     Colors encode the orientation of the vector fields, while the shading (from dark to bright) indicates their magnitude. Black dots mark points where the fields vanish.  
     (a) Original vector field $\tilde{v}_{\mathrm{1D}}(x)$ in real space and the corresponding order parameter space, i.e., the set of all orientations attained by $\vect{\tilde{v}}$, which in this example corresponds to a one-dimensional disk $\tilde{D}^1$.     
        (b) Resulting rescaled field $\vect{v}_\mathrm{1D}$ obtained via $\tilde{D}^1 \to {D}^1$. The associated order parameter space is the unit disc ${D}^1$.
        (c) Resulting normalized field $\vect{V}_\mathrm{1D}$ obtained through the embedding $\mathbb{R}^1 \to {S}^1$. Its associated order parameter space is the unit circle ${S}^1$. 
    }
    \label{fig:1to2}
\end{figure*}

\subsection{One-Dimensional Vector Fields}
\label{sec:1D}
We begin with the simplest case, namely one-dimensional scalar fields. Such fields occur in a wide range of physical systems, including domain walls in ferroic materials, solitons in conducting polymers, and superconducting order parameters near interfaces \cite{Mermin1979, Catalan2012, Su1979, DeGennes1966, Abrikosov1956}. In many of these examples, the field naturally attains zero values at defect cores or boundaries, making them particularly well suited to illustrate the embedding procedure.
Alternatively, the field may not become zero at the boundaries of a system, but rather take finite, non-vanishing values; this case is illustrated in App.~\ref{sec:DDW}.

As a representative example, we consider the scalar field
\begin{subequations}
    \begin{align}
        \tilde{v}_\mathrm{1D}&: \mathbb{R}^1 \rightarrow D^1 \subset \mathbb{R}^1,\\
        \tilde{v}_\mathrm{1D}(x) &= x e^{-x^2},
    \end{align}
\end{subequations}
which serves as a generic model for localized scalar textures and their associated gradients in condensed-matter and field-theoretic systems, see Fig.~\ref{fig:1to2}(a).
The field vanishes both at the origin and asymptotically as $x\rightarrow\pm\infty$, rendering a direct normalization ill-defined.
The image of the field is the interval
 $\tilde D^1=
\left[-1/\sqrt{2e},1/\sqrt{2e}\, \right]\subset \mathbb{R}^1$. 
The rescaling procedure described in Sec.~\ref{sec:Generalframework} therefore yields
\begin{subequations}
    \begin{align}
        v_\mathrm{1D}&:\mathbb{R}^1\rightarrow D^1\subset\mathbb{R}^1,\\
        v_\mathrm{1D}(x)&=\sqrt{2e}\, x e^{-x^2}.
    \end{align}
\end{subequations}
By construction, the image of $v_\mathrm{1D}$ coincides with the unit disk $D^1=[-1,1]$, see Fig.~\ref{fig:1to2}b).

The second step of the construction embeds the rescaled scalar field into one higher dimension. Applying Eq.~\eqref{eq:HigherMap} to $v_\mathrm{1D}$ yields the normalized vector field
\begin{subequations}
    \begin{align}
        \vect{V}_\mathrm{1D} &:\mathbb{R}^1 \rightarrow S^1 \subset \mathbb{R}^{2},\\
        \vect{V}_\mathrm{1D}(x)
            &= \begin{pmatrix}
                \sqrt{2e}\,x e^{-x^2} \\
                \Gamma_\mathrm{1D}(x) \, 
                \sqrt{1- (\sqrt{2e}\,x e^{-x^2})^2}
            \end{pmatrix}.
    \end{align}
\end{subequations}
Since $v_\mathrm{1D}(x)$ attains the values $\pm1$ at $x=\pm1/\sqrt{2}$, the second component vanishes at these points.
A continuous embedding is therefore obtained by choosing
\begin{equation}
    \Gamma_\mathrm{1D}(x) = \mathrm{sgn}\!\left(x^2-1/2\right),
\end{equation}
such that $\Gamma_\mathrm{1D}$ changes sign only where the square-root term vanishes. This choice ensures that $\vect{V}_\mathrm{1D}(x)$ remains continuous on the entire real line. 
The corresponding embedded field is shown in Fig.~\ref{fig:1to2}(c).
By construction, $|\vect{V}_\mathrm{1D}(x)|=1 $, and thus $\vect{V}_\mathrm{1D}$ defines a normalized mapping onto $S^1$. Since $\vect{V}_\mathrm{1D}(x)\to(0,1)^T$ for $x\to\pm\infty$, the real line can be compactified to $S^1$, yielding a continuous map
$\vect{V}_\mathrm{1D}^1\rightarrow S^1$,  classified by the homotopy group $\pi_1(S^1)=\mathbb Z$.
The corresponding winding number is obtained from
\begin{subequations}
    \begin{align} 
        Q_\mathrm{1D} &= -\frac{1}{2\pi} \int_{-\infty}^{\infty} \vect V_\mathrm{1D} \times \partial_x \vect V_\mathrm{1D} \,dx,\\
        &= \frac12 \left( \Gamma_\mathrm{1D}\left(\infty\right) - \Gamma_\mathrm{1D}\left(0\right) \right).
    \end{align}
\end{subequations}
For the choice of $\Gamma(x)$ defined above, the embedded field $\vect{V}_\mathrm{1D}$ has winding number $Q_\mathrm{1D}=1$.
Thus, the embedding procedure assigns a well-defined nontrivial topological invariant to the original scalar field despite the presence of amplitude zeros both at the origin and asymptotically at infinity.

\subsection{Two-Dimensional Vector Fields}
\label{sec:2D}

\begin{figure*}
    \includegraphics[width=\textwidth]{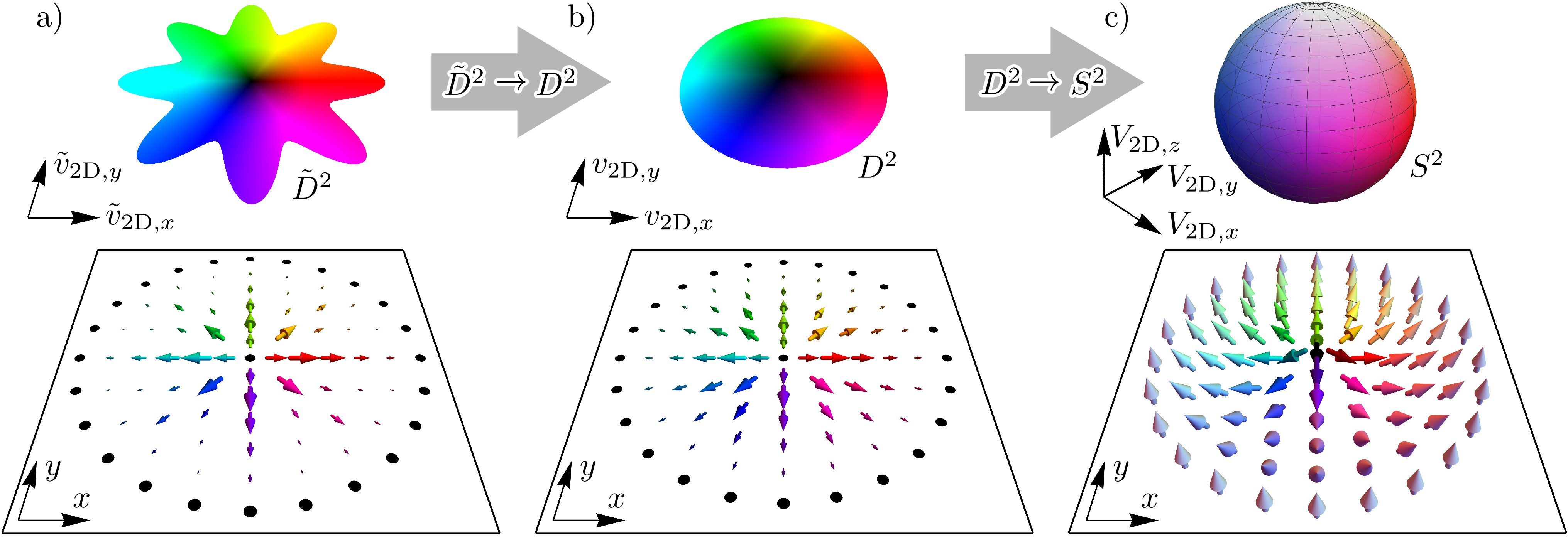}
    \caption{
    Two-dimensional example of the embedding procedure. (a) Original non-normalizable field $\tilde{\vect v}_{\mathrm{2D}}$ with flower-shaped target-space image $\tilde D^2$. (b) Rescaled field $\vect v_{\mathrm{2D}}$, whose image is the unit disk $D^2$. (c) Embedded normalized field $\vect V_{\mathrm{2D}}$, defining a mapping onto the unit sphere $S^2$. 
    Insets show the corresponding target-space images. Colors indicate the in-plane vector orientation, while the shading (from dark to bright) encodes the field magnitude in panels (a) and (b), and the out-of-plane component of $\vect V_{\mathrm{2D}}$ in panel (c). Black dots mark field zeros in 
    $\vect{\tilde{v}}_{\mathrm{2D}}$ and $\vect v_{\mathrm{2D}}$. The highlighted region in real space is mapped to the correspondingly highlighted region in target space.
}
\label{fig:2to3}
\end{figure*}

We now demonstrate the embedding procedure for two-dimensional vector fields. Physical systems in which the relevant observable is a two-dimensional vector field include in-plane magnetic textures, ferroelectric polarization fields, liquid-crystal order parameters, and velocity fields in fluid and wave systems \cite{Mermin1979, Hubert2008, Catalan2012, Shinjo2000, DeGennes1993, Nye1974}.

As an explicit example, we consider the non-normalizable but compactifiable vector field
\begin{subequations}
    \begin{align}
\vect{\tilde{v}}_{2D}&:\mathbb{R}^2 \rightarrow \tilde{D}^2 \subset \mathbb{R}^2,\\
    \tilde{\vect{v}}_{2D}(\vect{r}) &=  
    (1+1/4\cos(8\phi)) \, e^{1-r} \,\vect{r},
\label{eq:2to3NonNormal2D}
\end{align}
\end{subequations}
where $\vect r=(x,y)^T$, $r=\sqrt{x^2+y^2}$, and $\phi$ denotes the polar angle. 
This field describes a vortex texture whose magnitude vanishes at the origin and decays exponentially for $r\to\infty$. An eightfold modulation of the radial amplitude, introduced through the factor $1+\frac{1}{4}\cos(8\phi)$, gives rise to a target-space image that resembles an eight-petaled flower in $\mathbb{R}^2$. The corresponding real-space configuration and target-space image are shown in Fig.~\ref{fig:2to3}(a).

Following the general procedure introduced in Sec.~\ref{sec:Generalframework}, we first rescale $\tilde{\vect v}_{2D}$ to obtain a field $\vect v_{2D}$ whose image is the unit disk $D^2$,
\begin{subequations}
    \begin{align}
    \vect{v}_\mathrm{2D}&:\mathbb{R}^2 \rightarrow D^2 \subset \mathbb{R}^2,\\    
    \vect{v}_\mathrm{2D}
    & = \frac{\tilde{\vect{v}}_\mathrm{2D}}{R(\phi)}= e^{1-r}\, \vect{r},
\end{align}
\end{subequations}
where
$R(\phi)=1+1/4\cos(8\phi)$ 
denotes the maximal magnitude of $\tilde{\vect v}_{2D}$ for fixed angle $\phi$.
Fig.~\ref{fig:2to3}(b) shows the real-space configuration of $\vect v_\mathrm{2D}$ together with its corresponding order parameter representation.

In the second step, $\vect v_\mathrm{2D}$ is embedded into three dimensions. According to Eq.~\eqref{eq:HigherMap}, this yields
\begin{subequations}
\label{eq:HigherMap2D}
\begin{align}
    \vect{V}_\mathrm{2D} &:\mathbb{R}^2 \rightarrow S^2 \subset \mathbb{R}^{3},\\
    \vect{V}_\mathrm{2D}(\vect{r}) &=
\begin{pmatrix}
    \vect{r} e^{1-r} \\
    \Gamma_{\mathrm{2D}}(\vect r) \sqrt{1-(r e^{1-r})^2},
\end{pmatrix}
\end{align}
\end{subequations}
where $\Gamma_{\mathrm{2D}}(\vect r)=\mathrm{sgn} (\partial_r |\vect v_{2D}|)= \mathrm{sgn}\left(1-r\right)$ is chosen such that $\vect V_{2D}$ is continuous. By construction, $|\vect V_{2D}|=1$, such that $\vect V_{2D}$ defines a normalized mapping onto $S^2$. The embedded field and its order parameter image are shown in Fig.~\ref{fig:2to3}(c). In contrast to the original flower-like structure, the image now lies on the surface of the sphere $S^2$.

Together with the compactifiable base space, this yields a continuous mapping $\vect V_\mathrm{2D}: S^2 \rightarrow S^2$,
which can be classified by the homotopy group
$\pi_2(S^2) = \mathbb Z$.
The topological charge associated with the embedded field is
\begin{equation}
Q_\mathrm{2D}  = \frac{1}{4\pi}\int \mathrm{d}^2r \, \vect{V}_\mathrm{2D} \cdot \left( \partial_x \vect{V}_\mathrm{2D} \times \partial_y \vect{V}_\mathrm{2D} \right). \label{eq:topCharge2d}
\end{equation}
For the rotationally symmetric configuration shown in Fig.~\ref{fig:2to3}(c), this reduces to
\begin{equation}
Q_\mathrm{2D}= \frac{1}{2} \left. \left( \Gamma(\tilde{\vect v}(r)) \sqrt{1-|\tilde{\vect v}(r)|^2} \right) \right|_{0}^{\infty}.
\end{equation}
Since the field vanishes at $r=0$ and $r=\infty$, and the chosen $\Gamma_{\mathrm{2D}}(\vect r)$ satisfies $\Gamma(\tilde{\vect v}(\infty))=-1$ and $\Gamma(\tilde{\vect v}(0))=1$, we obtain $Q_\mathrm{2D}=-1$.
The embedded field therefore carries unit skyrmion charge and is classified as a skyrmion.

\begin{figure*}
    \centering
    \includegraphics[width=\textwidth]{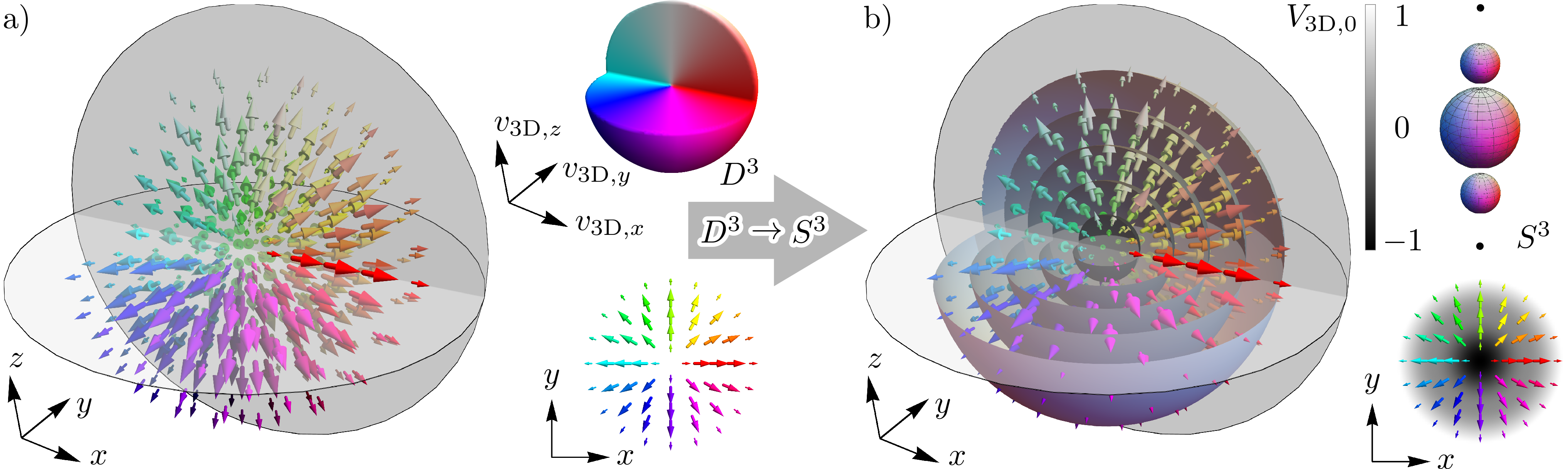}
    \caption{
        Three-dimensional example of the embedding procedure.
        (a) Illustration of the original non-normalizable vector field $\tilde{\vect{v}}_\mathrm{3D}(\vect{r})= \vect{v}_\mathrm{3D}(\vect{r})$. The lower inset shows the $xy$ cross section of the 3D structure. The upper inset shows the corresponding order parameter space $D^3$, which forms a filled sphere. 
        (b) Illustration of the embedded and normalized field $\vect{V}_\mathrm{3D}$. The $x$, $y$, and $z$ components of $\vect{V}_\mathrm{3D}$ are shown as color-coded vectors, following panel (a), and the fourth component $V_0$ is represented as an additional grayscale. For the chosen field, these isosurfaces form concentric shells as indicated in the figure. The lower inset shows the corresponding $xy$ cross-section. The four-dimensional order-parameter space $S^3$ is visualized through three-dimensional spheres with radii $|v_{\mathrm{3D},0}| = re^{1-r}$ in the upper inset. 
    }
    \label{fig:3to4}
\end{figure*}

\subsection{Three-Dimensional Vector Fields}
\label{sec:3D}

We now demonstrate the embedding procedure for three-dimensional vector fields. Examples of non-normalizable vector fields in three dimensions include electromagnetic fields in free space, ferroelectric polarization textures with vortex cores, and magnetic systems containing Bloch points or non-magnetic regions, where the field amplitude vanishes locally~\cite{DaCol2014, Naumov2004, Dennis2009, Azhar2022, Kuchkin2025, Kuchkin2026}.

The construction in three dimensions follows the general procedure outlined in Sec.~\ref{sec:Generalframework} and applied previously to the one- and two-dimensional examples.
Since the visualization of both the field and its target-space image becomes considerably more challenging in three dimensions, we consider as an explicit example a simple radially symmetric vector field whose image already coincides with the unit ball $D^3$. Consequently, $\tilde D^3=D^3$ and the rescaling step is trivial, i.e., $\tilde{\vect v}_{\mathrm{3D}}=\vect v_{\mathrm{3D}}$.
\begin{subequations}
\begin{align}
\tilde{\vect v}_{\mathrm{3D}}
&=
\vect v_{\mathrm{3D}}
:\mathbb{R}^3 \rightarrow D^3 \subset \mathbb{R}^3,
\\
\tilde{\vect v}_{\mathrm{3D}}(\vect r)
&=
\vect v_{\mathrm{3D}}(\vect r)
=
e^{1-r}\,\vect r, \label{eq.3dVec}
\end{align}
\end{subequations}
where $\vect r=(x,y,z)^T$ and $r=\sqrt{x^2+y^2+z^2}$. The field vanishes at the origin, reaches its maximal magnitude on the sphere $r=1$, and decays exponentially as $r\rightarrow\infty$, see  Fig.~\ref{fig:3to4}(a). The inset in Fig.~\ref{fig:3to4} shows the corresponding order parameter space, the three-dimensional unit disk $D^3$.
The second step consists of embedding $\vect v_{\mathrm{3D}}$ into four dimensions according to Eq.~\eqref{eq:HigherMap}, yielding
\begin{subequations}
\begin{align}
\vect V_{\mathrm{3D}}
&:\mathbb{R}^3\rightarrow S^3\subset\mathbb{R}^4,
\\
\vect V_{\mathrm{3D}}(\vect r)
&=
\begin{pmatrix}
e^{1-r} \vect{r}\\
\Gamma_{\mathrm{3D}}(\vect r)
\sqrt{1-(re^{1-r})^2}
\end{pmatrix}.
\end{align}
\end{subequations}
Since $r e^{1-r}\leq 1$, with equality only on the sphere $r=1$, a continuous embedding is obtained by choosing
$\Gamma_{\mathrm{3D}}(\vect r)
=\mathrm{sgn}(\partial_\vect{r}|\vect v_\mathrm{3D}|)
= \mathrm{sgn}(1-r).$
The resulting real-space vector field \(\vect{V}_\mathrm{3D}(\vect{r})\) is shown in Fig.~\ref{fig:3to4}(b), together with the associated \(S^3\) order parameter space next to it. 
Because the four-dimensional field cannot be visualized directly, we show its $x$, $y$, and $z$ components as color-coded vectors, following panel (a), and represent $V_{\mathrm{3D},0}$ by grayscale isosurfaces. For the chosen field, these isosurfaces form concentric shells. The four-dimensional order-parameter space is therefore depicted as three-dimensional spheres with radii $|v_{\mathrm{3D},0}| = re^{1-r}$.
By construction, $|\vect V_{\mathrm{3D}}|=1$, and the compactification of the base space is preserved. 
Consequently, $\vect V_{\mathrm{3D}}$ defines a normalized continuous map $S^3\rightarrow S^3$, whose topology is classified by $\pi_3(S^3)=\mathbb Z$. The corresponding topological invariant is
\begin{equation}
Q_\mathrm{3D} = \int \frac{\mathrm{d}^3r}{2\pi^2} \,\epsilon_{\alpha\beta\gamma\delta} V_\mathrm{3D,\alpha}(\partial_x\, V_\mathrm{3D,\beta})(\partial_y\, V_\mathrm{3D,\gamma})(\partial_z\, V_\mathrm{3D,\delta}),
\end{equation}
where $\alpha,\beta,\gamma,\delta\in[x,y,z,0]$ and $\epsilon_{\alpha\beta\gamma\delta}$ is the four-dimensional Levi–Civita symbol with $\epsilon_{xyz0}=1$. 
For the choice of $\Gamma_{\mathrm{3D}}$ introduced above, the image of $\vect V_{\mathrm{3D}}$ covers the target sphere exactly once, yielding the topological charge
$Q_{\mathrm{3D}}= \frac{1}{2}\left(\Gamma_\mathrm{3D}(\infty)-\Gamma_\mathrm{3D}(0)\right)=-1$.

\section{Discussion \& Conclusion}
\label{sec:discon}
We have introduced a general framework for assigning topological invariants to non-normalizable vector fields. The central idea is to continuously deform the target-space image of a field into a unit disk and subsequently embed the resulting field into one higher dimension, yielding a normalized mapping onto a sphere. Whenever the base space can be compactified, the embedded field defines a continuous map between spheres and can therefore be classified using the corresponding homotopy group.

The construction was illustrated for representative one-, two-, and three-dimensional vector fields. In each case, the procedure yields a well-defined topological invariant despite the presence of amplitude zeros that would preclude a conventional normalization. These examples demonstrate that non-normalizable fields can nevertheless possess robust topological structure when viewed through the higher-dimensional embedding introduced here. An important aspect of the construction is the freedom in the choice of the lifting function $\Gamma$, which determines how the original field is embedded into the higher-dimensional sphere. 
The topological invariants of the embedded field inherit the robustness associated with homotopy theory. As a consequence, the topological charge assigned to the original non-normalizable field cannot change under continuous deformations that preserve the embedding. A transition between distinct topological sectors is only possible when the assumptions underlying the construction break down, for example, through the creation or annihilation of field zeros or a change in the topology of the target-space image.

The framework developed here considerably broadens the range of vector fields that can be analyzed using topological methods. It applies to a variety of physical systems, including electromagnetic fields, ferroic order parameters, and magnetic textures containing amplitude zeros or singular regions. 
An important direction for future work is to elucidate the role of the associated topological invariants in the underlying physical systems and to determine whether they give rise to measurable signatures, particularly, stabilized states, or novel dynamical phenomena.

\section{Acknowledgments}
We thank Finn Boyer, Pascal Dreher, Volodymyr Kravchuk, Sandra Chulliparambil Shaju and Martin Speight for discussions. We acknowledge funding from the German Research Foundation (DFG) 
Project No.~278162697-SFB 1242 (project B10 and project B06), 
Project No.~403233384 (SPP2137 Skyrmionics),
Project No.~405553726-CRC/TRR 270 (project B12),
Project No.~505561633 in the TOROID project co-funded by the French National Research Agency ANR under Contract No. ANR-22-CE92-0032.
M.A.\ acknowledges support from the UDE Postdoc Seed Funding Project TORUS.

\appendix

\section{Choice of Lift} 
\label{sec:choiceOfLift}
As discussed in Sec.~\ref{sec:Generalframework}, the choice of the scalar function $\Gamma(\vect v)$ is generally not unique. This nonuniqueness has a geometric origin. For every point $\vect v$ in the interior of the unit disk $D^n$, the embedding construction associates two possible points on the sphere $S^n$, corresponding to the two hemispheres distinguished by the sign of the additional component. The function $\Gamma$ determines which of these two points is selected and therefore specifies how the image of $\vect v$ is lifted from $D^n$ to $S^n$.
The two possible lifts coincide only on the boundary of the disk, where $|\vect v|=1$ and the additional component vanishes. Consequently, continuity permits sign changes of $\Gamma$ only across the preimage of the boundary $\partial D^n$. Different continuous choices of $\Gamma$
therefore correspond to different ways of assigning regions of the image $D^n$ to the northern and southern hemispheres of $S^n$. A particularly simple choice is $
\Gamma(\vect v)\equiv 1$, which maps the entire image of $\vect v$ to a single hemisphere. Since a
hemisphere is contractible, the resulting map is homotopically trivial and therefore carries vanishing topological charge. In contrast,
nontrivial topological sectors arise when $\Gamma$ changes sign across the image, such that the embedded field covers both hemispheres of the target sphere.
For compactifiable vector fields, the embedding defines a continuous map
\begin{equation}
\vect V : S^m \rightarrow S^n,
\end{equation}
whose topology is classified by the homotopy group $\pi_m(S^n)$.
Different choices of $\Gamma$ may therefore lead to different homotopy
classes of the embedded field. In the examples discussed in the main text, the trivial lift yields vanishing topological charge, while the
two simplest nontrivial lifts correspond to topological charges of opposite sign. More generally, the freedom in choosing $\Gamma$ reflects a geometric freedom in lifting a map from $D^n$ to $S^n$, with distinct lifts corresponding to distinct topological sectors of the embedded field.
It should be emphasized that the trivial lift always exists and is independent of the detailed structure of the original vector field. Consequently, it does not encode information about the topology of the underlying configuration and is therefore of limited physical interest. The nontrivial lifts, by contrast, capture global information contained in the original field and give rise to the topological invariants considered in this work.

\section{Example with a Nonzero Asymptotic Field}
\label{sec:DDW}

\begin{figure*}[tb]
    \centering
    \includegraphics[width=\textwidth]{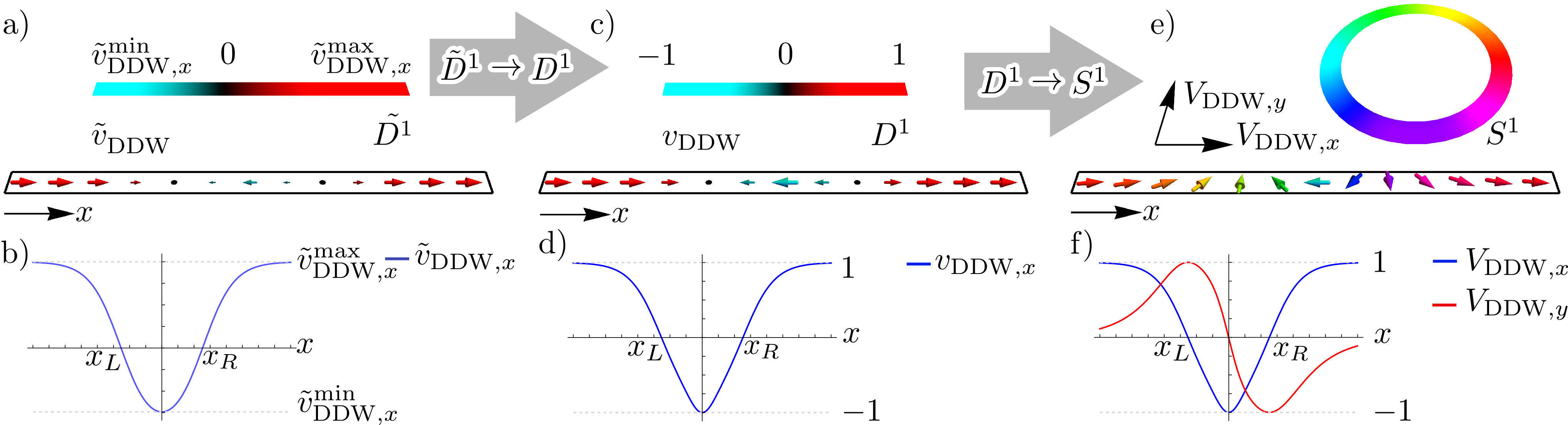}
    \caption {
    One-dimensional double-domain-wall example of the radial-rescaling and embedding procedure.
    (a) Original non-normalizable scalar field $\widetilde{v}_{\mathrm{DDW}}$ in real space, together with its target-space image $\widetilde{D}^{\,1}$. The target-space interval extends from the minimum value $\widetilde{v}^{\min}_\mathrm{DDW}$ to the maximum value $\widetilde{v}^{\max}_\mathrm{DDW}$.
    (b) Corresponding double-domain-wall profile, with $x_{\mathrm{L}}$ and $x_{\mathrm{R}}$ indicating the positions of the left and right domain walls, respectively.
    (c) Radially rescaled field $v_{\mathrm{DDW}}$, whose target-space image is the unit interval $D^1=[-1,1]$.
    (d) Corresponding real-space profile of the rescaled field.
    (e) Normalized field $\mathbf{V}_{\mathrm{DDW}}$ obtained by lifting $v_{\mathrm{DDW}}$ onto the unit circle $S^1$. The inset shows the resulting target-space image on $S^1$.
    (f) Cartesian components $V_{\mathrm{DDW,x}}$ and $V_{\mathrm{DDW,y}}$ of the embedded field.
    In panels (a) and (c), the color indicates the value of the scalar field, while in panel (e) it encodes the orientation of $\mathbf{V}_{\mathrm{DDW}}$ on $S^1$. 
    }
\label{fig:DDW}
\end{figure*}

As discussed in the main text, the mapping procedure applies naturally to compactifiable fields whose amplitude vanishes at infinity.
A second relevant class is formed by fields that approach a uniform, nonzero value at infinity.
In this case, the boundary of real space can still be identified, such that the base space is compact.

As an illustrative example, we consider the one-dimensional double-domain-wall (DDW) profile
\begin{equation}
    \tilde{v}_{\mathrm{DDW}}(x) = \tanh\!\left(\frac{x-x_{\mathrm{R}}}{w_{\mathrm{R}}}\right) - \tanh\!\left(\frac{x-x_{\mathrm{L}}}{w_{\mathrm{L}}}\right) + 1 .
    \label{eq:DDW_profile}
\end{equation}
Here, $x_{\mathrm{L}}$ and $x_{\mathrm{R}}$ denote the positions of the left and right domain walls, respectively, while $w_{\mathrm{L}}$ and $w_{\mathrm{R}}$ control the corresponding domain-wall widths.
For $x\to\pm\infty$, the field approaches the uniform value $\tilde{v}_{\mathrm{DDW}}\to 1$ so that the compactification of the real line is well defined.
The image of this scalar field is a compact, generally asymmetric one-dimensional disk $  \tilde{D}^1= [\tilde{v}^{\min}_\mathrm{DDW},\tilde{v}^{\max}_\mathrm{DDW}] \subset \mathbb{R}$, where $\tilde{v}^{\min}_\mathrm{DDW}$ and $\tilde{v}^{\max}_\mathrm{DDW}$ are the minimal and maximal values attained by $\tilde{v}_{\mathrm{DDW}}$.
This target-space interval is shown in Fig.~\ref{fig:DDW}(a), while the corresponding real-space profile is shown in Fig.~\ref{fig:DDW}(b).
Before applying the higher-dimensional lifting procedure, the field must first be rescaled from $\tilde{D}^1$ to the unit interval $D^1=[-1,1]$, as described in Sec.~\ref{sec:Generalframework}. For the profile shown here, this rescaling is implemented by introducing the smooth positive scaling function
\begin{equation}
    f(x) = \frac{1}{\left(|\tilde{v}^{\min}_\mathrm{DDW}|-|\tilde{v}^{\max}_\mathrm{DDW}|\right)e^{-x^2} + |\tilde{v}^{\max}_\mathrm{DDW}|},
\end{equation}
and defining
\begin{equation}
    v_{\mathrm{DDW}}(x) = \tilde{v}_{\mathrm{DDW}}(x) f(x).
\end{equation}
The resulting field $v_{\mathrm{DDW}}$ is shown in Fig.~\ref{fig:DDW}(c), with the corresponding real-space profile shown in Fig.~\ref{fig:DDW}(d).
The final step is the lift to the unit circle,
\begin{equation}
    \vect{V}_{\mathrm{DDW}}(x)
    =
    \begin{pmatrix}
        v_{\mathrm{DDW}}(x) \\
        \Gamma_{\mathrm{DDW}}(x)
        \sqrt{1-v_{\mathrm{DDW}}^2(x)}
    \end{pmatrix}
    \in S^1 ,
\end{equation}
where $\Gamma_{\mathrm{DDW}}(x)=\mathrm{sgn}(\partial_x v_\mathrm{DDW}(x))$ is chosen such that $\vect{V}_{\mathrm{DDW}}$ remains continuous. 
For the nontrivial lift shown in Fig.~\ref{fig:DDW}(e) and (f), the image of $\vect{V}_{\mathrm{DDW}}$ winds once around $S^1$. The corresponding winding number is therefore
$Q_{\mathrm{DDW}} = 1$. 
This example demonstrates that the mapping procedure is not restricted to fields that decay to zero at infinity. It can also be applied to compactifiable structures with a uniform nonzero boundary value and an asymmetric target-space image. The same concept extends directly to higher-dimensional compactifiable fields.

\newpage

\bibliography{Refs}
\end{document}